\documentclass[10pt,a4paper]{article}
\usepackage[latin1]{inputenc}
\usepackage{amsmath}
\usepackage{amsfonts}
\usepackage{amssymb}
\usepackage{makeidx}
\usepackage{graphicx}
\usepackage{todonotes}
\usepackage{lipsum}
\usepackage{tabularx}
\usepackage{float}
\usepackage{multirow}

\providecommand{\keywords}[1]
{
	\small	
	\textbf{Keywords---} #1
}

\author{
	Rik van Leeuwen$^{1}$ $\bullet$ Ger Koole$^{2}$ \\
	\texttt{\scriptsize\sffamily $^{1}$Ireckonu, Olympisch Stadion 43, 1076DE, Amsterdam The Netherlands} \\
	\texttt{\scriptsize\sffamily $^{2}$Department of Mathematics, Vrije Universiteit, De Boelelaan 1111, 1081HV Amsterdam, The Netherlands} \\
	\texttt{\scriptsize\sffamily $^{1}$rik@ireckonu.com $\bullet$ $^{2}$ger.koole@vu.nl} \\
}
\title{Demand Forecasting in Hospitality Using Smoothed Demand Curves }
\date{\today}

\begin{document}
	
	\maketitle
	
	\begin{abstract}
	Demand forecasting is one of the fundamental components of a successful revenue management system. This paper provides a new model, which is inspired by cubic smoothing splines, resulting in smooth demand curves per rate class over time until the check-in date.This model makes a trade-off between the forecasting error and the smoothness of the fit, and is therefore able to capture natural guest behavior. The model is tested on hospitality data. We also implemented an optimization module, and computed the expected improvement using our forecast and the optimal pricing policy. Using data of four properties from a major hotel chain, between 2.9 and 10.2\% more revenue is obtained than using the heuristic pricing done by the hotels. 
	\end{abstract}

	\keywords{Revenue Management, Forecasting, Cubic Smoothing Splines}
	
	\vspace{1cm}
	
	\section{Introduction}
	\label{sec: intro}
	
	In hospitality, revenue management (RM) systems typically consist of a forecasting model and a decision support model, where the success of the system heavily depends on the performance of both models and the interaction between them (Rajopadhye et al., 1999). Based on historical reservations, a forecast is created which serves as input for decision optimization that determines the rate given a capacity of the perishable product. Therefore, forecasting demand is one of the key inputs for a profitable RM system (Weatherford \& Kimes, 2003). In the strongly related airline industry, a 10\% increase in forecast accuracy increases revenue by 0.5-3.0\% on high demand flights (Lee, 1990).
	
	To forecast demand, research has been conducted on different types of models such as time series and machine learning techniques (Claveria et al., 2015). The models typically incorporate features such as seasonality or demand fluctuations to increase the forecast accuracy. Whereas research has predominantly focused on the forecast accuracy to compare the models, there has been less focus on the interpretability and explainability. In the hospitality industry, it is seen as important (Andrew et al., 1990). 
	
	Medium-sized and larger hotel chains have dedicated teams of revenue managers who determine the rates for future dates. They often make their decisions based on gut feeling and information on, for example,  \textit{booking pace} and occupancy. Booking pace is defined as the speed at which reservations are made for a check-in date. To move towards more data-driven decision-making in the hospitality industry, a RM system can support revenue managers. As the success of a system highly depends on the willingness of revenue managers to adopt, it is important for a system to be transparent, also known as a 'white-box' system (Loyola-González, 2019)).
	
	Transparency can be created by sharing demand curves with revenue managers. A demand curve represents the demand over time until a given check-in date, hereafter referred to as demand scenario. Demand requests are typically used to create the demand scenarios. As requests are constrained by the capacity of the property and the rate of a room at a specific point in time, unconstrained optimisation methods are typically used to forecast demand (Queenan, 2009).  The benefit of unconstrained optimisation methods for RM systems is between $0.5-1.0\%$ in terms of revenue growth (Weatherford \& Polt, 2002).
	
	The development of demand for a demand scenario is dynamic; e.g., demand usually increases over time towards a check-in date (Haensel \& Koole, 2010). To capture these dynamics, splines are often applied as unconstraining optimisation to forecast demand because they are sufficiently flexible. A spline is a piecewise polynomial function, which can extrapolate and is therefore a suitable technique. This technique is used in other applications as well; e.g., in water demand models (Santopietro et al., 2020) and cost transportation models (Rich, 2018). The essence of the cubic splines model is applied in this paper, which is based on a third-degree polynomial as interpolant. 
	
	This paper proposes a white-box forecasting framework that takes into account the practical side of RM. The cubic smoothing spline model has never been used in the hospitality industry and can also be used on other industries such as the airline and car rental industry. The framework is created based on an extensive data analysis in Section~\ref{sec:data analysis}, after a description of the data in Section~\ref{sec: data source}. The forecast model is presented in Section \ref{sec: model}, followed by the decision support model in Section~\ref{sec: decision}. The accuracy measurements are given in Section~\ref{sec: AccuracyMeasurement}. The model is tested in a controlled environment in Section~\ref{sec:simulation}. The empirical results are discussed in Section~\ref{sec:results} and the conclusion and discussion are presented in Section~\ref{sec:conclusion}. 

	\section{Data}
	\label{sec: data source}
	The data is pulled from a Property Management System (PMS), a system used in the hospitality industry to register and regulate reservations. Table~\ref{tab:reservation} shows the attributes attached to a reservation. Person-related attributes are not included because of the General Data Protection Regulation (GDPR). Only relevant attributes are shown in the table. The hotel that is used for this research has a single room type and therefore room type is not included.
	
	The rate of a reservation is the total amount, however, the rate per night may differ if the length of stay is larger than one. An additional table is joined to the reservation table, where rates for individual nights are stored. This one-to-many relationship between the reservations table and rates table provides individual night rates. For this research, multi-stay reservations are transformed into individual stay nights. The individual night rates are used as demand requests to create the demand scenarios. 
	
	\newpage
	
	Reservations with a rate equal to 0 are excluded. These are part of a complimentary service where guests stay without a charge. These guests are mostly part of the hospitality chain itself or part of marketing campaigns (e.g., social media influencers).
	
	\begin{table}[h]
		\centering
		\begin{tabularx}{\columnwidth}{X|X}
			\hline
			\textbf{Variable} 	& \textbf{Description} \\ \hline
			Arrival date 		& Date of arrival of reservation   \\
			Departure date 		& Date of departure of reservation   \\
			Length of stay 		& Number of days between arrival date and departure date  \\
			Booking date 		& Date that the reservations was made \\
			Status				& Status of reservation, either 'stay', 'cancellation' or 'no-show' \\
			Rate 				& Total amount in euro \\
			Group				& Boolean indicating part of group reservation \\
			Source 				& Main channel reservation  \\
			Sub source 			& Specified channel of reservation \\
			Market code 		& Statistical information on booking and revenue development   \\ \hline
		\end{tabularx}
		\caption{Reservation attributes and descriptions}
		\label{tab:reservation}
	\end{table}

	Data of a four properties, located in the Netherlands, United Kingdom and, United States of America, are included from the 1st of January 2017 until the 31st of December 2019. With three years of data, yearly seasonality and trends over time can be captured. The names and locations of the properties are anonymized. Each property has a minimum and maximum room night rate that a revenue manager can set, these settings are displayed in Table \ref{tab:DataHotelRateSettings}.  
	
	\begin{table}[h]
		\begin{tabularx}{\columnwidth}{X|XXX}
                 & \multicolumn{3}{c}{\textbf{Rate Settings}}          \\ \cline{2-4} 
			& \textbf{Minimum} & \textbf{Maximum} & \textbf{Step} \\ \hline
			\textbf{Hotel 1} & 70               & 170              & 10            \\
			\textbf{Hotel 2} & 90               & 240              & 15            \\
			\textbf{Hotel 3} & 100              & 250              & 15            \\
			\textbf{Hotel 4} & 150              & 450              & 20    
		\end{tabularx}
			\caption{Rate settings by property}
			\label{tab:DataHotelRateSettings}
	\end{table}

	\section{Data Analysis}
	\label{sec:data analysis}
	The following key performance indicators (KPIs) are used in the hospitality industry to analyse the performance of a property according to Pizam (2010) and Mauri (2012). Average Daily Rate (ADR) is the sum of room rates from all the checked-out reservations divided by the number of occupied rooms. Revenue Per Available Room (RevPAR) is the sum of room rates from stayed reservations in the hotel divided by the capacity. Occupancy (Occ) is the ratio between the number of occupied rooms and the capacity. Although some rooms are \textit{out-of-order} or \textit{complementary}, the KPIs in this paper are based on the full capacity of the property. The results of the data analysis is shown only for Hotel 1 since results of the other properties are inline with Hotel 1.
	
	\newpage
	
	Reservations can only have a single status: stay, canceled, or no-show. On average, $74.2\%$ of the nights are marked as stay, $23.3\%$ of the nights as canceled, and $2.5\%$ of the nights as no-show. Table~\ref{tab:status} shows the reservations by status per year in percentages. In 2019, the cancellation rate is slightly lower compared to $2017$ and $2018$. The no-show rate is consistent over the years.
	
	\begin{table}[h]
		\centering
		\begin{tabularx}{\columnwidth}{X|XXX}
								& \textbf{Stay} & \textbf{Cancellation} & \textbf{No-show} 	\\ \hline
			2017     			& 72.97        	& 24.30        			& 2.72           	\\
			2018     			& 72.82        	& 24.84        			& 2.34           	\\
			2019     			& 77.02        	& 20.67        			& 2.31           	\\ \hline
			\textbf{Overall} 	& 74.22        	& 23.33        			& 2.46          
		\end{tabularx}
		\caption{Reservation status by year in percentages}
		\label{tab:status}
	\end{table}

	The ADR by status per year is shown in Table~\ref{tab:statusADR}. Over the years, the ADR increased, indicating the room rate increased over time. ADR for canceled and no-show also increased over the years. RevPAR for 2017, 2018, and 2019 is $71.9$, $73.8$ and $79.6$, respectively. RevPAR only includes reservations with status stay.
	
	\begin{table}[h]
		\centering
		\begin{tabularx}{\columnwidth}{X|XXX|X}
								& \textbf{Stay} & \textbf{Cancellation} & \textbf{No-show} 	& \textbf{Overall}	\\ \hline
			2017     			& 98.5  		& 104.8 				& 103.0 			& 100.2 			\\
			2018     			& 101.4 		& 103.3 				& 103.7 			& 101.9 			\\
			2019     			& 103.4 		& 106.4 				& 105.6 			& 104.1 			\\ \hline
			\textbf{Overall} 	& 101.1 		& 104.7 				& 104.0	    		& 102.0
		\end{tabularx}
		\caption{ADR per year by reservation status}
		\label{tab:statusADR}
	\end{table}
	
	\subsection{Seasonality}
	The hospitality industry is characterised by seasonality throughout the months and days stated by Shields (2013). When seasonality is poorly understood, it may lead to a slow pace of sales or even lost sales. In this section, an analysis is presented of the seasonality by month and day.
	
	\subsubsection{By Month}
	\label{subsec: DA - Month}
	 Table~\ref{tab:DataAnalyticsMonth} shows the three KPIs by month over the available dataset. Differences over the months are consistent throughout the year for each KPI. Unlike resort properties in popular vacation destinations (Chu, 2009), this property doesn't experience any KPI fluctuations by month. 
	 
	\begin{table}[h]
		\centering
		\begin{tabularx}{\columnwidth}{X|XXX}
			& \textbf{ADR} & \textbf{RevPAR} & \textbf{Occ} (in \%) \\ \hline
			\textbf{Jan} & 101.0       & 82.8           & 82.30      \\
			\textbf{Feb} & 101.1       & 82.0           & 82.12      \\
			\textbf{Mar} & 101.3       & 81.3           & 80.88      \\
			\textbf{Apr} & 101.3       & 82.1           & 81.28      \\
			\textbf{May} & 101.2       & 82.6           & 81.50      \\
			\textbf{Jun} & 101.2       & 82.3           & 80.87      \\
			\textbf{Jul} & 101.3       & 80.9           & 80.23      \\
			\textbf{Aug} & 100.9       & 80.7           & 80.08      \\
			\textbf{Sep} & 100.9       & 81.4           & 80.81      \\
			\textbf{Oct} & 101.1       & 84.8           & 82.74      \\
			\textbf{Nov} & 100.9       & 84.2           & 82.37      \\
			\textbf{Dec} & 101.0       & 82.5           & 81.52     
		\end{tabularx}
		\caption{KPIs per month - ADR and RevPar in local currency}
		\label{tab:DataAnalyticsMonth}
	\end{table}
	
	\subsubsection{By Day}
	\label{subsec: DA - DoW}
	Table~\ref{tab:dow} shows the three KPIs by day over the available dataset. The property typically attracts business guests during the week and leisure guests during the weekend. On Friday and Saturday, the occupancy is highest. On Sunday, the occupancy is lowest because leisure guests depart on Sunday and business guests arrive on Monday. Because of lower demand on Sunday, the ADR is lowest.
	
	\begin{table}[h]
		\centering
		\begin{tabularx}{\columnwidth}{X|XXX}
							& \textbf{ADR}   & \textbf{RevPAR} 	& \textbf{Occ}  (in \%)	 \\ \hline
		\textbf{Monday}    	& 100.2 		 & 80.7   			& 80.53\%        \\
		\textbf{Tuesday}   	& 104.3 		 & 89.8  			& 86.08\%        \\
		\textbf{Wednesday} 	& 106.2 		 & 91.9   			& 86.56\%        \\
		\textbf{Thursday}  	& 102.7 		 & 86.3  			& 83.98\%        \\
		\textbf{Friday}    	& 97.4  		 & 87.0  			& 89.33\%        \\
		\textbf{Saturday}  	& 108.3 		 & 101.8			& 93.96\%        \\
		\textbf{Sunday}    	& 84.8  		 & 49.2  			& 58.00\%           
		\end{tabularx}
		\caption{KPI's by day of week from 2017 until 2019}
		\label{tab:dow}
	\end{table}

	\newpage

\begin{figure}[h]
		\centering
		\includegraphics[width=1\linewidth]{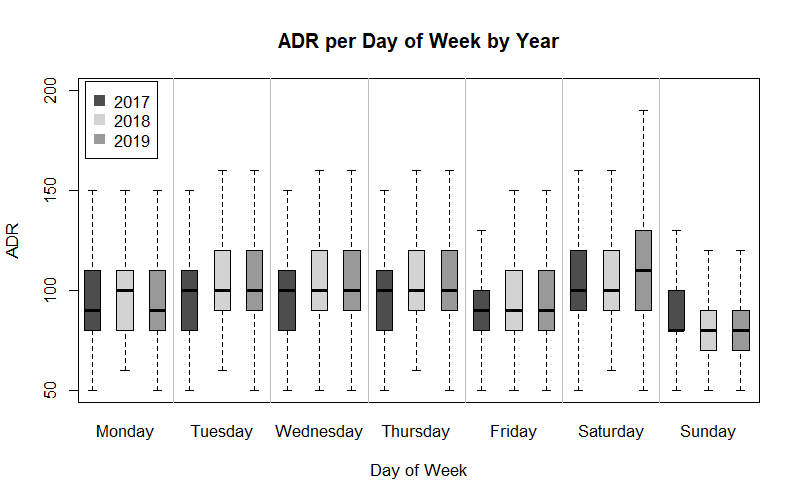}
		\caption{Metric ADR per day of week by year}
		\label{fig:DA - ADR DoW Year}
	\end{figure}

	A detail view of ADR by day is displayed in Figure~\ref{fig:DA - ADR DoW Year}. Except Sunday, there is an increase in ADR over the years. This trend is present for every day, except Sundays, where there is a decrease in ADR over the years. The highest variability is on Saturday in 2019. 

\begin{figure}[h]
		\centering
		\includegraphics[width=1\linewidth]{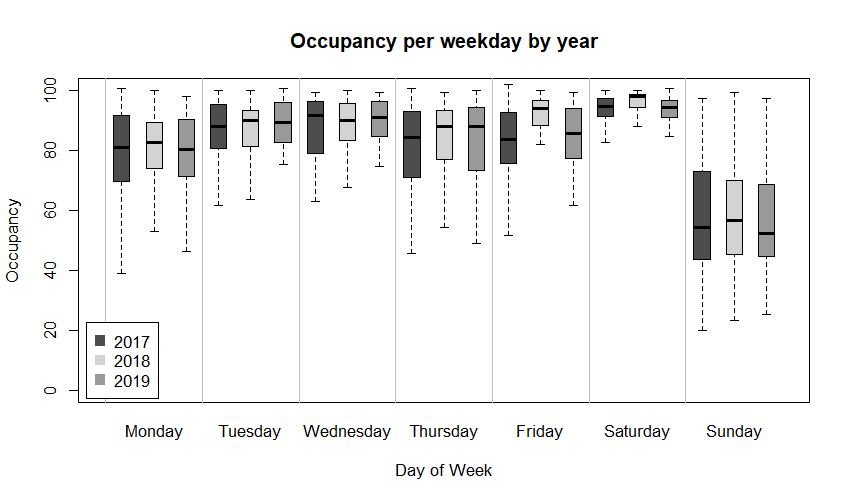}
		\caption{Metric occupancy per day of week by year}
		\label{fig:DA - Occ DoW Year}
	\end{figure}

	A detailed view of occupancy by day is displayed in Figure~\ref{fig:DA - Occ DoW Year}. The range of occupancy by day varies over the years. Overall, 2018 varied less in occupancy compared to 2017 and 2019. 
	
	For seasonality patterns by day, the relationship between the booking day of the reservation and the day of the stay is often analysed. This feature is engineered because it is not part of the original dataset. Table~\ref{tab:DA - DoW Arrival / Booking} shows the relationship between the day the reservation has been made and the day the guest spent a night at the property. For example, guests who stay on a Monday reserve the most on Mondays ($19.78$\%) and the least on Saturdays ($6.5$\%). There is more fluctuation from Monday to Thursday compared to Friday to Sunday. Business travelers tend to book more during weekdays and less during weekend days. For weekend days, guests tend to book more constantly throughout the week. 
	
	\begin{table}[h]
		\centering
		\begin{tabular}{l|lllllll}
		 & \textbf{Mon} & \textbf{Tue} & \textbf{Wed} & \textbf{Thu} & \textbf{Fri} & \textbf{Sat} & \textbf{Sun} \\ \hline
		\textbf{Mon}    & 19.78   & 15.82     & 16.11    & 16.26  & 16.39    & 6.5    & 9.14  \\
		\textbf{Tues}   & 19.4    & 20.09     & 15.71    & 16.02  & 16.43    & 5.17   & 7.18  \\
		\textbf{Wed} 	& 17.89   & 20.11     & 19.69    & 15.73  & 14.68    & 5.06   & 6.84  \\
		\textbf{Thu}  	& 16.69   & 17.78     & 19.11    & 18.75  & 13.74    & 5.94   & 7.99  \\
		\textbf{Fri}    & 15.52   & 15.71     & 15.31    & 15.8   & 16.04    & 9.04   & 12.57 \\
		\textbf{Sat}  	& 14.8    & 15.11     & 14.82    & 14.6   & 14.46    & 13.02  & 13.2  \\
		\textbf{Sun}    & 14.22   & 15.08     & 15.5     & 15.35  & 15.04    & 10.31  & 14.49
		\end{tabular}
		\caption{Distribution of reservations per stay day of week (rows) made on booking day of week in percentages}
		\label{tab:DA - DoW Arrival / Booking}
	\end{table}
	
	\newpage
	
	\subsection{Lead Time}
	\label{subsec: DA - LeadTime}
	Lead time is the number of days between the booking date and arrival date. This variable is not present in the dataset, and is therefore engineered. Lead time $0$ is the last day a guest can make the reservation. The maximum lead time is the check-in date, which is $365$ days in the available dataset. This range implies reservations can be made one year in advance. For analysis purposes, the booking horizon is used as the range of making a reservation, where 0 is the first day and 365 as the last day  to make a reservation. By applying the booking horizon, time is progressing over an x-axis. By analysing this feature, more knowledge is gained when guests make reservations, which is important in demand forecasting.

\begin{figure}[H]
		\centering
		\includegraphics[width=.5\textwidth]{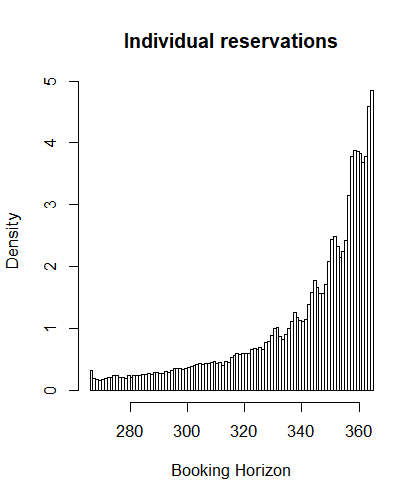}
		\caption{Density for lead time 265 - 365, left individual and right cumulative }
		\label{fig:DA - Lead Time density}
	\end{figure}

	Only $1$\% of reservations are made in the range $0$ - $146$ days before the check-in date. Around $5$\% of the reservations are booked between $0 - 252$ days. This implies that $95$\% of the reservations are made in the last 100 days before check-in date. In the last 7 days before the check-in date, $27$\% of the reservations are made. In Figure \ref{fig:DA - Lead Time density}, the density over the last 100 days before the check-in date is shown. The cumulative density shows a smooth gradient towards $100$\%.
	
	\newpage

	Reservations made at $t$ of the booking horizon for stays on specific day of the week (see Section \ref{subsec: DA - DoW}), the density of the booking horizon is further analysed by day in the Appendix \ref{sec: Appendix - Lead Time density by DoW} Figure~\ref{fig:DA - Lead Time density by DoW}. The density doesn't drop at the same time $t$, indicating that the underlying demand until the check-in day is influenced by booking day. For example, for Tuesdays, fewer reservations are made in weekends, whereas for Saturdays, the density is more granular over the booking horizon.

	\section{Model}
	\label{sec: model}
	A new method is proposed to forecast the demand in hospitality, which estimates a smooth unconstrained curve. A demand curve is defined as the number of guests who are willing to buy a product over a period of time. In this paper, a product is a room for a specific rate. The booking horizon is defined from the first day a room can be reserved until check-in date $T$, where $t=1,...,T$. The advantage of this method is that the unconstrained demand curve is estimated without any assumptions on the shape by applying ideas similar to cubic smoothing splines.
	
	\subsection{Cubic Splines}
	A spline is a piecewise polynomial function that only contains non-negative integer powers of $x$. The polynomials can be determined by spline interpolation, where the interpolant is a third-degree polynomial. Spline interpolation is preferred to polynomial interpolation because it avoids the problem of Runge's Phenomenon, where oscillation occurs at the edges of the interval (Epperson, 1987). A spline $S(x)$ function is defined for a set of $n$ observations \(\{x_{k}, y_{k}: k = 1,...,n\}\), where a single $y$ value exists for each $x$:	
	
	\begin{enumerate}
		\item $S: \mathbb{R} \rightarrow \mathbb{R}$ 
		\item $S(x)$ is a polynomial of degree 3 of each subinterval \([x_{k}, x_{k+1}]\) with $k=1,...,n-1$.
		\item $S(x_{k})=y_{k}$ $\forall$ $k=1,...,n$. 
	\end{enumerate}
	
	The spline $S(x)$ is a combination of $i$, for $i=1,...,n-1$, polynomials of degree 3, which is defined as:
	
	\begin{equation}\label{spline}
		S(x) = 
			\begin{cases}
				C_{1}(x),  & \text{for }x_{1}\le x<x_{2}\\
				... \\   
				C_{i}(x), & \text{for }x_{k}\le x<x_{k+1}\\ 
				...\\
			    C_{n-1}(x), & \text{for }x_{n-1}\le x<x_{n}
			\end{cases}
	\end{equation}
	
	Subjected to several conditions, the system can be solved and so the spline function can be defined for a given set of observations. These conditions are listed below:
	
	\begin{itemize}
		\item $C_{i}(x_{i})=y_{i}$ and $C_{i}(x_{i+1})=y_{i+1}$ $\forall$ \(i=1,...,n-1\)
		\item \(C'_{i}(x_i) = C'_{i+1}(x_{i}) \) $\forall$ \(i=1,...,n-2\)
		\item \(C''_{i}(x_i) = C''_{i+1}(x_{i}) \) $\forall$ \(i=1,...,n-2\)
		\item \(C'_{1}(x_1) = 0 \) and \(C'_{n-1}(x_{n})=0 \) 
	\end{itemize}
	
	The last condition is known as the natural boundary condition, which prevents the spline to oscillate. The number of coefficients to be estimated is $4(n-1)$, because \(C_{i}=a_{i}+b_{i}x+c_{i}x^{2}+d_{i}x^3\) for $i=1,...,n-1$. This form of interpolation is known as cubic splines because the polynomial is a cubic.

	This method can be applied in hospitality where $\hat{S}$ represents the demand curve for a specific demand scenario. Value $x$ represents the number of days of the booking horizon and value $y$ represents the number of reservations. 
	
	\subsection{Cubic Smoothing Splines}
	When there are multiple $y$ values, indicated by the vector $Y$, per $x$, the optimization problem changes, since the spline cannot be defined in each point of $y$ for a given $x$. Therefore, a smoothed spline is required. Given a set of $n$ observations \(\{x_{k}, Y_{k}: k = 1,...,K\}\), modelled by the relation \(Y_{k} = S(x_{k}) + \varepsilon_{k}\) where \(\varepsilon_{k}\) is independent and identically distributed.
	
	The cubic smoothing spline estimate $\hat{S}$ of the function $S$, where $g$ represents the smoothing parameter, is defined as
	
	\begin{equation}\label{smoothingspline}
		\min \quad (1-g)\sum_{k}^{} \{\hat{S_k} - Y_{k} \}^{2} + g \int_{}^{} \hat{S}^{''}(x)^{2}dx
	\end{equation}
	
	In Equation~(\ref{smoothingspline}), the first term measures the closeness to the data and the second term penalizes the curvature in the function. The smoothing parameter, $g$, controls the balance between these terms and should be larger than~0 and smaller than~1. If $g = 0$, it indicates no smoothing and the result function is completely focussed on the data. If $g = 1$, it indicates infinite smoothing and the result is a straight function. 
	
	This model is limited due to the inability of adding constraints that represents industry knowledge, i.e., demand cannot drop below zero and dependencies between multiple rates.
	
	\subsection{Linear Transformation}
	The essence of the cubic smoothing spline model is transformed into a linear function to avoid quadratic minimization. Therefore, both terms of the minimization function are transformed into linear terms. The first term is transformed by taking the absolute value between the data point and fit. In doing so, the function becomes more robust against outliers. The second term is transformed to the second difference, $d^{(2)}$. The first difference is defined as $d_{k} = S(x_{k-1}) - S(x_{k})$. As long as the second derivative exists and is continuous, the second difference is defined. The smoothing parameter controls the balance between the two terms and should be larger or equal than~0 and smaller or equal than~1. The linear minimization function is defined as follows:.
	
	\begin{equation}\label{eq:smoothingsplinelinearfunction}
		\min \quad (1- g)\sum_{k}^{} {|e_{k}| \cdot w_{k}} + g \sum_{k=3}^{n}{|d^{(2)}_{k}|}
	\end{equation}

	The $w_{k}$ variable represents the weight of each distinct $y$ value to reduce the number of constraints. The sum of this variable is equal to 1, so the weight per $x$ value is set to 1 divided by the number of $y$ values in $k$. The polynomial is defined as \(C_{i}=a_{i}+b_{i}x+c_{i}x^{2}+d_{i}x^3\) for $i=1,...,n-1$.

	 By transforming it into a linear minimization function, this problem can be solved by adding constraints to a Linear Programming (LP) model. Since many LP programs assume that all variables are non-negative, an extra variable per decision variable, $e$ and $d^{(2)}$, is added to model the absolute values. 

	\begin{itemize}
		\item $e_{k} \geq S(x_{k}) - Y_{k} $ $ \forall$ \(k=1,...,n\)
		\item $e_{k} \geq -(S(x_{k}) - Y_{k})$ $ \forall$ \(k=1,...,n\)
		\item $d^{(2)}_{k} \geq d_{k-1} - d_{k}$ $ \forall$ \(k=3,...,n\)
		\item $d^{(2)}_{k} \geq -(d_{k-1} - d_{k})$ $ \forall$ \(k=3,...,n\)
		\item \(C_{k}(x_k) = C_{k+1}(x_{k}) \) $ \forall$ \(k=1,...,n-1\)
		\item \(C'_{k}(x_k) = C'_{k+1}(x_{k}) \) $ \forall$ \(k=2,...,n-1\)
		\item \(C''_{k}(x_k) = C''_{k+1}(x_{k}) \)$ \forall$ \(k=2,...,n-1\)
		\item \(C'_{1}(x_1) = 0 \) and \(C'_{n-1}(x_{n})=0\)
		\item \(e_k\) $\in$ $\mathbb{R}^{+}$ $\forall$ \(k=1,...,n\)
		\item \(d_k\) $\in$ $\mathbb{R}^{+}$ $\forall$ \(k=3,...,n\)
		\item \(a_i, b_i, c_i, d_i\) $\in$ $\mathbb{R}$ $ \forall$ \(i=1,...,n-1\)
	\end{itemize}
	
	An additional constraint that is required is the demand to be larger than~$0$ over the booking horizon. If this restriction is not added, it implies that rooms are given away for free. This constraint is modeled as follows:
	\begin{itemize}
		\item $S(x_k) \geq 0$ $ \forall$ \(k=1,...,n\)
	\end{itemize}

	At least three distinct $x$-values are required to derive the second difference for the curvature penalty. 
	
	The number of polynomials, $C$, is equal to the distinct $x$-values minus 1. Each polynomial contains 4 coefficients. The error between the data and the fit is denoted by $e$. The number of $e$ decision variables is equal to the number of distinct $x$ values. The number of decision variables to model the curvature penalty $d$ is $(n - 2)$. To model unrestricted decision variables in an LP model, an additional decision variable is required (dependent on the implementation). In total, the number of decision variables in the LP model is equal to $(4 \times (n - 1) + n + (n - 2)) \times 2$. For example, when the input includes 4 weeks, or 28 days, the total number of decision variables is equal to 324.
	
	\subsubsection{Rates}
	\label{sec: customerchoisesets}
	Observations with multiple $y$ values for a single $x$ can indicate arrivals for different rates on a specific day. As multiple rates for a single room are common in the hospitality industry, the cubic spline model is adjusted. Rates are denoted as $r$, for $r=1,...,R$. 
	
	Given $N$ observations \(\{x_{r, k}, Y_{r, k}: k = 1,...,n ; r:1,...,R\}\) where $Y_{r, k}$ represents the number of sales for a given rate on a given day on the booking horizon. The spline $S_{r}$ now represents the demand curve for a rate $r$: 
	
	\begin{equation}\label{smoothingsplinelinearfunctionrates}
		\min \quad (1-g_{r})\sum_{r}^{}\sum_{k}^{} {|e_{r, k}| \cdot w_{r, k}} + g_{r} \sum_{r}^{} \sum_{k=3}^{n} {|d^{(2)}_{r, k}|}
	\end{equation}
	
	An additional constraint is added to the LP model to incorporate that people are always willing to book a room if the rate of a room is lower than their willingness-to-pay (Haensel \& Koole, 2010). For example, when a room can be booked for a rate of 300 then that person was also willing to book that room for a rate of 200 or even 100. The rates of $100$, $200$, and $300$ are defined as A, B, and C, respectively. Rate A represents the lowest rate for a room. These typical choice sets are defined as the set of alternatives in order of preference. For the example, choice set {A, B, C} imply that guests are willing to spend more than 100, choice set B, C imply that guests are willing to spend more than 200, and choice set C imply that guests that are willing to spend 300 or more. Given choice sets are strictly ordered, an additional constraint is added to the linear model: 
	
	\begin{itemize}
		\item $S_{r} \geq S_{r+1}$ $ \forall$ \(r=1,...,R-1\)
	\end{itemize}

	The total number of decision variables in the LP model where rates are included is equal to $(4 \times (n - 1) + n + (n - 2)) \times 2 \times r$. 
	
	\newpage

	\subsection{Residuals}
	To assess the goodness-of-fit, the residuals are analysed. The residuals are the difference between the actual values and estimated values. For each $t$ of the booking horizon, a transformation to the input values is applied to ensure the residuals are independent of demand.
	
	Arrivals follow an inhomogeneous Poisson process with a rate of $\lambda_t$ (Lee, 1990). Consequently, it is expected that the standard deviation of the residuals increases (and thus the goodness of fit decreases) when the rate increases. To stabilize the standard deviation, the square root of the original data points is taken as input to the model (Brown et al., 2001). This transformation is known as Anscrombe transform.

	\section{Decision Optimization}
	\label{sec: decision}
	The performance of the forecasting model can be expressed in revenue by applying decision optimization. The performance is expressed in revenue since that is the main driver for the hospitality industry. Dynamic programming (DP) is a widely studied and adapted technique for decision optimization (Talluri~\&~Ryzin~(2004a) and Talluri~\&~Ryzin~(2004b)). DP uses demand estimates in order to maximize revenue given the capacity of a property. This section provides a brief description of the application of DP in hospitality.
	
	As multiple bookings are not considered per $t$ on the booking horizon to ensure rate changes after each sold item, the time is divided into small intervals such that only one booking can be made in an interval. The variable $r_j$ denotes a rate, for $j=1,...,R$. The different classes are ordered as follows: $r_1 > r_2 > ... > r_j > ... > r_R$. The probability of a booking for rate $r_{j}$ at a time $t$ is denoted as $\lambda(r_{j}, t)$. Parameter $\lambda(r_{j}, t)$ assumes an inhomogeneous Poisson process, where $\lambda(r_{j}, t) < 1$.
	
	The system is modeled as a mathematical function of random disturbances and control decisions according to Bellman's equation. The objective is to find a control policy that maximizes the total expected revenues by solving the value function $V$, which is defined as:  
	
	\begin{equation}
		\label{bellman}
		V_{t}(x) =  \max_{j \in\{1,...,R\}} \{\, \lambda(r_{j}, t) \cdot r_{j} + V_{t+1}(x - 1) + (1-\lambda(r_{j}, t)) \cdot V_{t+1}(x) \,\}  \,
	\end{equation}
	
	By following Bellman's principle for optimality, the value $V_{t}(x)$ denotes the total expected value at time $t$, where time $t=1$ represents the day a room is available for a booking and $t=T$ the check-in date, given capacity $x$.  The total expected revenue is obtained by calculating $V_{1}(x)$, given a capacity $x$.
	
	When solving the value function $V$, two boundary conditions exist. Firstly, the value function becomes 0 for any t when no capacity is left (that is, $V_{t}(0)=0$). Secondly, the value function becomes $0$ for any $T$ when there is no time left to make a reservation (that is, $V_{T}(x)=0$). 
	
	\section{Accuracy Measurement}
	\label{sec: AccuracyMeasurement}
	Accuracy is measured between the observed value $a$ and the forecasted value $f$. Observations $a$, used for fitting a model, are referred as in-sample data points. Where observations $a$, used to measure accuracy, are known as out-of-sample data points. A distinction needs to be made between results that use in-sample data points and out-of-sample data points.  
	
	To measure the accuracy of the model, the Weighted Absolute Percentage Error (WAPE) is applied:
	
	\begin{equation}\label{wape}
		WAPE = \frac{\sum_{t=1}^{n}|a_t - f_t|}{\sum_{t=1}^{n} a_t}
	\end{equation}
	
	Amongst the accuracy measurements, WAPE is the most appropriate metric since it takes into account the actuals by weight. This metric sets importance on higher values of actual values. Unlike, for example Mean Percentage Error (MPE) and Mean Absolute Percentage Error (MAPE), WAPE is defined when actual demand is zero at time $t$. The Mean Absolute Error (MAE) and Mean Squared Error (MSE) are depended on the capacity of the property itself, which blocks the ability to compare properties with each other. 
	
	The result indicates the closeness between the observed values $a$ and forecasted values $f$. Demand is assumed to follow a Poisson distribution. This results in a baseline error due to the variability of the distribution.

	Given the demand curves over time for each choice set, the decision support model optimises the expected revenue given the capacity left on that $t$ of the booking horizon. The revenue of the actual result of a demand scenario is compared to the optimal expected revenue, and expressed in percent change. To summarize, performance is expressed in two ways, WAPE and the revenue percent change. The WAPE is used to express the fit of the demand curve to the data points and revenue percent change indicates the performance of the model from a business perspective.
	
	\section{Simulation}
	\label{sec:simulation}
\begin{figure}[H]
		\centering
		\includegraphics[width=1\linewidth]{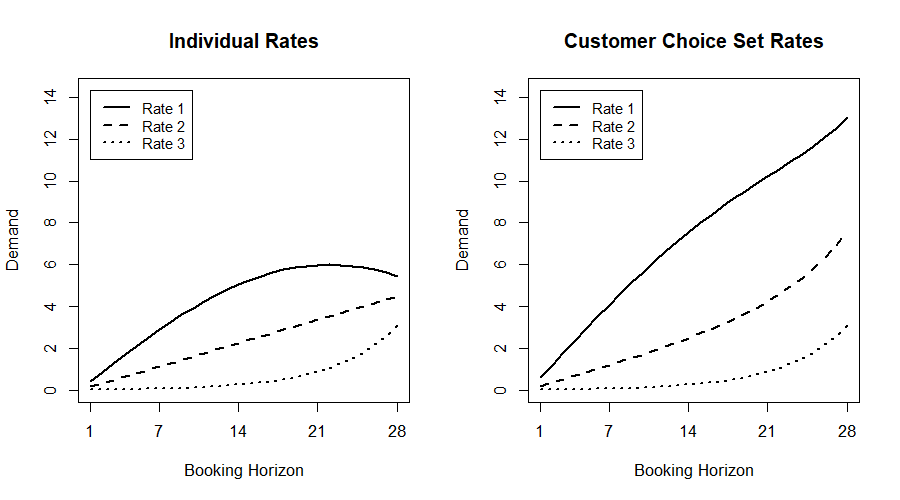}
		\caption{Simulation scenario with demand curves (L) and customer choice set demand curves (R)}
		\label{fig:simulationscenario}
	\end{figure}

	To test the validity of the cubic smoothing spline model, simulations are performed using sets of three known 'individual rate' classes (see Figure~\ref{fig:simulationscenario}). Following the definition of rates in Section~\ref{sec: customerchoisesets}, the 'customer choice set rates' are created. That is, choice set rate class~$2$ is the sum of individual rate class~$1$~and~$2$ and choice set rate class~$3$ is the sum of individual rate class~$1$,~$2$ and~$3$. Guests are always willing to spend less for the same type of room, and thus the demand is added. In the simulations, the booking horizon is set to $28$ days, where $t=28$ is the check-in date.
	
	Table~\ref{tab:Simulation - Rate Class Parameters} shows the functions used for the three 'individual rates'. Rate class~$1$ represents guests who are looking for the best deal. Close to the end of the booking horizon, demand decreases due to increasing pressure from competition. Rate class 3 represents business guests who book a room last minute and only make reservations near the check-in date. Rate class 2 is a combination of the two previously mentioned guests and is simulated as constant demand. 
	
	\begin{table}[H]
		\centering
		\begin{tabularx}{\columnwidth}{X|X}
			\textbf{Rate} & \textbf{Curve Parameters} \\ \hline
			\textbf{1}    & $0.43 * \sin(t)$	 \\
			\textbf{2}    & $0.16 * t $ \\
			\textbf{3}    & $0.02 * \exp(.18 * t)$             
		\end{tabularx}
		\caption{Parameters of rate classes used for simulation}
		\label{tab:Simulation - Rate Class Parameters}
	\end{table}
	
	Demand at time $t$, denoted as $\lambda_t$, is simulated by setting a single rate for a given $t$ of the booking horizon. A rate is open by sampling from a trial, where each rate class has a chance to be open. Finally, a Poison distribution with rate $\lambda_t$ is applied to perform a simulation of arrivals. Cancellations or no-shows are not included in the simulations, implying data isn't generated when maximum capacity is reached. 
	
	To calculate the expected revenue, a price is assigned to each rate class. The three rates classes and their corresponding prices are defined as $r=\{100, 200, 300\}$. The expected optimal revenue is $17,327$ given a capacity of $100$ rooms, which has an ADR of $173.3$. 
	
\begin{figure}[H]
		\centering
		\includegraphics[width=1\linewidth]{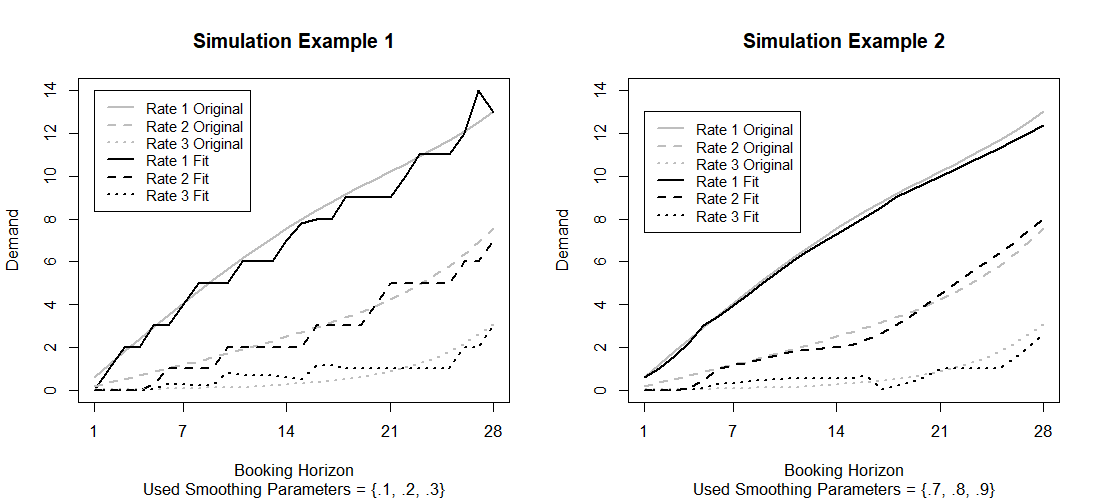}
		\caption{Cubic smoothing spline model based on 50 simulated demand scenarios with two different smoothing parameter settings}
		\label{fig:simulationexample}
	\end{figure}

	The cubic smoothing spline model is fitted to 50 demand scenarios using two different sets of smoothing parameters. The left-hand-side of Figure~\ref{fig:simulationexample} is the result with smoothing parameters \{$.1, .2, .3$\} and the right-hand-side is the result with smoothing parameters \{$.7, .8, .9$\}. The optimal expected revenue with smoothing parameters \{$.1, .2, .3$\} and \{$.7, .8, .9$\} are equal to $16,819$ and $17,020$, respectively. This is equal to an ADR of $168.2$ and $170.2$, respectively.
	
\begin{figure}[H]
		\centering
		\includegraphics[width=.7\linewidth]{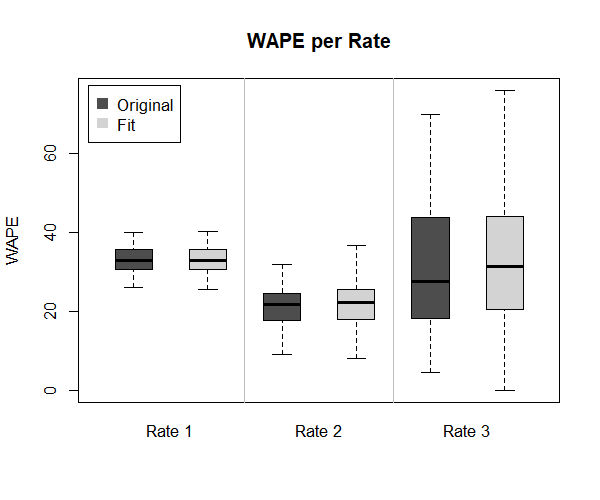}
		\caption{Out-of-sample results based on individual demand scenarios expressed in WAPE for original demand curves and estimated demand curves with smoothing parameter settings \{.7, .8, .9\}}
		\label{fig:simulationoutsample}
	\end{figure}

	The original and fitted demand curves are shown in Figure~\ref{fig:simulationexample}, where the black lines are the fitted curves and the grey lines the original curves. The WAPE per rate class is $7.16$, $,11.5$, and $81.0$ with smoothing parameter settings \{$.1, .2, .3$\} and $6.11$, $12.26$ and $44.07$ with smoothing parameter settings \{$.7, .8, .9$\}. These results can be marked as in-sample of the model. A new set of 100 demand scenarios are simulated from the original demand curves to analyze out-of-sample results. The newly simulated demand scenarios are compared to the original demand scenarios and the fitted demand scenarios by the model. Figure~\ref{fig:simulationoutsample} shows per rate class two box plots, where the WAPE with the original demand curves as fitted values is compared to the WAPE with the demand curves from the model as fitted values. The mean as well as the variation are comparable, indicating that the LP model is very precise.
	
	\subsection{Sensitivity Analysis}
	To test the robustness of the cubic smoothing spline model, the relationship between the input and output is analyzed. These simulations use the demand curves presented in Figure~\ref{fig:simulationscenario}. In the simulation setting, two input parameters can be changed: (1) the number of demand scenarios and (2) the smoothing parameter. The output of the model is expressed in WAPE and revenue (see Section~\ref{sec: AccuracyMeasurement}).
	
\begin{figure}[H]
		\centering
		\includegraphics[width=1\linewidth]{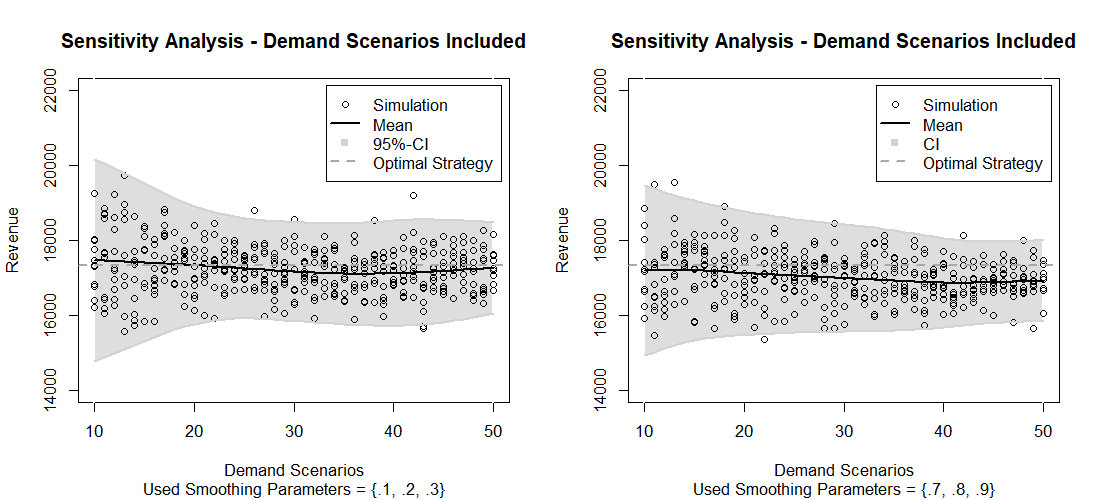}
		\caption{Sensitivity analysis regarding number of input demand scenarios with respect to revenue with two different smoothing parameter settings}
		\label{fig:simulationsensitivityscenarios}
	\end{figure}

	The number of demand scenarios is one of the input parameters that can be controlled. As there is always a trade-off between increase in accuracy and increase in computational time, the simulation setup includes 10 to 50 demand scenarios to understand the effect of the number of demand scenarios. The setup is executed 10 times. In total, 410 runs are executed. The smoothing parameters are fixed for both smoothing parameters settings \{$.1, .2, .3$\} and \{$.7, .8, .9$\}. Figure~\ref{fig:simulationsensitivityscenarios} shows the sensitivity of the number of demand scenarios, with the number of demand scenarios on the x-axis and the revenue per simulation set-up on the y-axis. The mean is steady over the range of demand scenarios for both smoothing parameter settings. The speed at which the confidence interval narrows decreases slowly when more demand scenarios are included. Therefore, including more demand scenarios lead to a stable result when the demand curves remain equal. 
		
	\section{Results}
	\label{sec:results}
	In this section, the results our model from the empirical data are discussed. The complete dataset, containing data from 2017 to 2019, is used as input. First, the results are presented. Second, the set-up of the model is explained. And finally, the performance of the model is assessed.
	
	To provide additional insights about the model and the application of it, Thursdays of Hotel 1 are taken as use-case. To create the customer choice set, the rates from every Thursday in $2018$ are included. The rates range between $70$ and $170$. A class is defined as $10$ units, and thus $11$ rate classes are present in the data. Table~\ref{tab:ResultThursdayExample} shows the number of reservations made in the last $28$~days before the check-in date. The customer choice set, which is input for the model, is defined following the definition in Section~\ref{sec: customerchoisesets}.
	
	\begin{table}[H]
		\centering
		\begin{tabular}{l|ll}
			\textbf{Rate} & \textbf{Reservations made} & \textbf{Customer Choice Set logic} \\ \hline
			\textbf{70}   & 201                  & 2508                               \\
			\textbf{80}   & 316                  & 2192                               \\
			\textbf{90}   & 481                  & 1711                               \\
			\textbf{100}  & 449                  & 1262                               \\
			\textbf{110}  & 362                  & 900                                \\
			\textbf{120}  & 256                  & 644                                \\
			\textbf{130}  & 246                  & 398                                \\
			\textbf{140}  & 126                  & 272                                \\
			\textbf{150}  & 87                   & 185                                \\
			\textbf{160}  & 55                   & 130                                \\
			\textbf{170}  & 75                   & 75                                
		\end{tabular}
		\caption{Number of reservations for every Thursday in 2018}
		\label{tab:ResultThursdayExample}
	\end{table}
	
	As 2018 counts $52$ Thursdays, $52$ demand scenarios are included. Therefore, the smoothing parameters are set lower. If more data is available, the natural behavior of demand patterns can be best captured by lowering the smoothing parameter. The range is between $0.1$ for Rate~$70$ and $0.5$ for Rate~$170$. Using linear interpolation, the other parameters are set between $0.1$ and $0.5$. 
	
\begin{figure}[H]
		\centering
		\includegraphics[width=.5\linewidth]{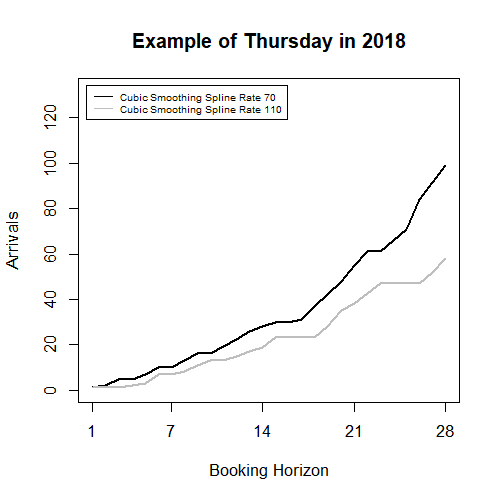}
		\caption{Results of model - demand curves for Rate 70 and Rate 110}
		\label{fig:ResultExampleThursdays}
	\end{figure}

	Figure~\ref{fig:ResultExampleThursdays} shows the demand curves from the cubic smoothing spline model with rate 70 (black line) and rate 110 (gray line). The results are in line with the analysis in Section~\ref{sec:data analysis}. The demand is increasing towards the check-in date and the demand is decreasing during days in the weekend (i.e., the curves flatten at $t=16,17$ and $t=23,24$). 

	\subsection{Setup}
	To set up the cubic smoothing spline model, dates from 2017 and 2018 are used to forecast a date in 2019. Dates are only included in the input if the day is smaller as the day that is forecasted. If a Tuesday is forecasted by the model, every Tuesday until the forecasted date is used as input. The booking horizon is 100 days. 
	
	The last 28 days before check-in are forecasted for each date in 2019. The first 72 days of the booking horizon are used to fit the model. The optimal expected revenue of the first 72 days is compared with historical dates. WAPE is used to measure the accuracy of the model over the first 72 days. A pitfall for only considering the revenue at $t=1,...,72$ could be that the behavior of two demand scenarios could deviate extremely.
	
\begin{figure}[H]
		\centering
		\includegraphics[width=1\linewidth]{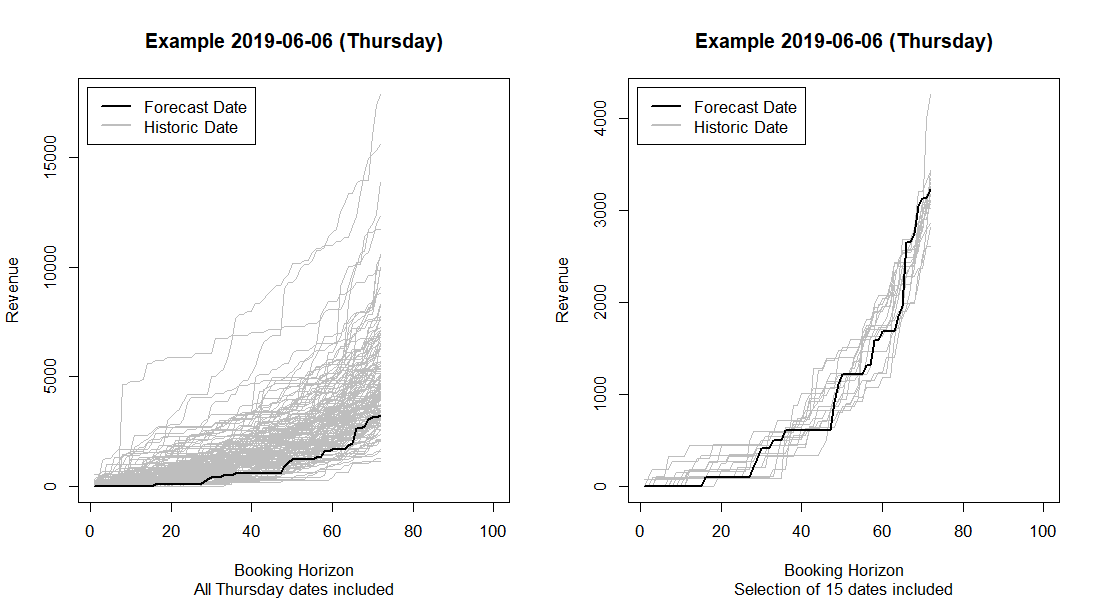}
		\caption{Example of dates selection which serves as input for the model}
		\label{fig:resultsdatesselection}
	\end{figure}

	In total, 15 dates are selected as input for the cubic smoothing spline model. Figure~\ref{fig:resultsdatesselection} shows the revenue of Thursday the 6th of June 2019, a randomly chosen date. The left-hand side shows the revenue for all days in 2017 and 2018. The right-hand side shows the dates that are used as input for the model (only Thursdays). Note that the scales of the y-axes in Figure~\ref{fig:resultsdatesselection} are different. In both figures, the black line represents the revenue on the 6th of June 2019. The WAPE including all days raging between $17.35$ and $746.85$, with a mean of $116.40$. The WAPE including the dates used as input for the model raging between $17.35$ and $27.07$, with a mean of $22.57$. By taking into account the 15 closest days, measured in WAPE, contributes positively to selection of demand scenarios as input for the model.
	
	Based on the 15 dates as input, demand curves are generated per rate class. These demand curves per rate are used as input for the decision optimization algorithm to obtain the optimal expected revenue. The number of reservations made for the last 28 days of the forecasted data are used as the capacity available. The actual revenue is compared with the optimal expected revenue and expressed in percent difference. The WAPE between the actual bookings and demand curves are generated to gather data about the closeness of the fit and actual data points. The smoothing parameters remain consistent among results, which is linear over 11 rates starting from $0.4$ and which a maximum of $0.7$.
	
	\subsection{Performance}
	The performance of the model is measured in revenue between the actual revenue and the optimal expected revenue, and the WAPE of the fitted demand scenarios. The rate range differs per property, based on Table \ref{tab:DataHotelRateSettings} in Section~\ref{sec: data source}. The results are presented by hotel by day of the week. 
	
	For each of the properties, the model generates on average between 2.9\% - 10.2\% more revenue. The details, expressed in mean and standard deviation, per property and day of week are presented in Table~\ref{tab:resultperformancewapemeanhotel}. Given all day of weeks of the properties, 85.7\% in a positive result. For Tuesday and Wednesday for Hotel 1 and Hotel 2, the model did not outperform the actual revenue. Since the standard deviation is larger than the mean, not every forecasted demand scenario ended up in a positive result. However, overall the forecasting model generates more revenue then hospitality practitioners, e.g., revenue managers. 
	
	\begin{table}[H]
		\centering
		\begin{tabular}{ll|llll}
			&               & \textbf{Hotel 1} & \textbf{Hotel 2} & \textbf{Hotel 3} & \textbf{Hotel 4} \\ \hline
			\multicolumn{1}{l|}{\multirow{2}{*}{\textbf{Monday}}}    & \textbf{Mean} & 4.24             & 3.93             & 13.66            & 12.61            \\
			\multicolumn{1}{l|}{}                                    & \textbf{SD}   & 15.45            & 15.95            & 16.88            & 20.34            \\ \hline
			\multicolumn{1}{l|}{\multirow{2}{*}{\textbf{Tuesday}}}   & \textbf{Mean} & -0.15            & -5.71            & 5.87             & 7.49             \\
			\multicolumn{1}{l|}{}                                    & \textbf{SD}   & 16.36            & 12.48            & 19.74            & 18.65            \\ \hline
			\multicolumn{1}{l|}{\multirow{2}{*}{\textbf{Wednesday}}} & \textbf{Mean} & -2.57            & -3.51            & 7.22             & 6.35             \\
			\multicolumn{1}{l|}{}                                    & \textbf{SD}   & 12.26            & 13.83            & 21.22            & 19.93            \\ \hline
			\multicolumn{1}{l|}{\multirow{2}{*}{\textbf{Thursday}}}  & \textbf{Mean} & 7.53             & 4.54             & 12.79            & 12.31            \\
			\multicolumn{1}{l|}{}                                    & \textbf{SD}   & 18.84            & 18.59            & 19.34            & 19.31            \\ \hline
			\multicolumn{1}{l|}{\multirow{2}{*}{\textbf{Friday}}}    & \textbf{Mean} & 4.48             & 6.99             & 13.32            & 9.13             \\
			\multicolumn{1}{l|}{}                                    & \textbf{SD}   & 15.6             & 16.05            & 18.7             & 17.73            \\ \hline
			\multicolumn{1}{l|}{\multirow{2}{*}{\textbf{Saturday}}}  & \textbf{Mean} & 0.09             & 11.32            & 9.42             & 4.61             \\
			\multicolumn{1}{l|}{}                                    & \textbf{SD}   & 15.91            & 17.28            & 17.99            & 15.13            \\ \hline
			\multicolumn{1}{l|}{\multirow{2}{*}{\textbf{Sunday}}}    & \textbf{Mean} & 6.14             & 1.49             & 8.54             & 9.27             \\
			\multicolumn{1}{l|}{}                                    & \textbf{SD}   & 10.51            & 11.04            & 15.69            & 14.9             \\ \hline
			\multicolumn{1}{l|}{\multirow{2}{*}{\textbf{Overall}}}   & \textbf{Mean} & 2.92             & 3.25             & 10.19            & 8.89             \\
			\multicolumn{1}{l|}{}                                    & \textbf{SD}   & 15.46            & 16.05            & 18.57            & 17.95           
		\end{tabular}
		\caption{Mean and standard deviation of WAPE by hotel by day of week}
		\label{tab:resultperformancewapemeanhotel}
	\end{table}

	The dates selection in order to fit the model is executed first, which is based on 15 demand scenarios that are the closest in terms of revenue over the booking horizon $t=1,...,72$. This closeness is measured in WAPE over the booking horizon. Appendix~\ref{sec: Appendix - Results WAPE} contains four tables that report the WAPE per property per day of week per customer choice set. In general, the mean WAPE is lower for lower rates and higher for higher rates. This is because historically lower rates are set more often. Therefore, more data points are present in the available set for lower rates than higher rates. In general, there is an increasing trend towards higher rates, however this trend decays toward the highest rates. The variability amongst rates increases from the lowest towards the highest rates. These findings are in line with the results of the simulation, where an increasing trend is noticed when less data is available.

	\section{Conclusion and Discussion}
	\label{sec:conclusion}
	This paper proposes a novel method inspired by cubic smoothing splines, which makes use of linear programming. As input, the model uses reservations for specific rates over time. As output, it provides demand curves per rate over a given booking horizon, demonstrating the booking behavior of guests.  
	
	The data analysis shows that demand does not differ significantly throughout the years 2017, 2018, and 2019. However, booking behavior differs among the day of the week. These differences are expressed in standard hospitality KPI's such as ADR, RevPAR, and occupancy. The day of booking, with respect to the check-in date, influences the demand significantly. The booking horizon is a variable that is essential for the proposed forecasting framework. There is a clear distinction when guests tend to make a reservation for a given day of week.
	
	The performance of the model is evaluated based on the fit to the data points as well as optimal expected revenue. The model claims to generate $2.9\% - 10.2\%$ more revenue based on the forecasted demand curves of each rate. The revenue difference is significant but realistic since all the rates are set manually by a revenue management team. In terms of WAPE, the model shows similar values as the simulation environment for lower rates. When moving towards higher rates, the WAPE tends to increase. This can be caused by less available data on higher rates.
	
	Hospitality practitioners, e.g., revenue managers, can directly benefit from such a white-box method by analysing the demand curve per customer choice set until check-in date. It enables the revenue managers to understand how demand is progressing over time. The transparency of this method will help practitioner' willingness to adopt a revenue management system to increase revenue on a daily basis.
	
	More research could be conducted into the smoothing parameters. During the simulations and empirical study, this parameter was set regardless of the rate. When more data points are available, the smoothing parameter needs to be set lower to better capture the natural behavior of demand. Based on the available data, the smoothing parameters can be further optimised by setting them for each rate independently.
	
	More research can also be conducted into the number of fitted polynomials. The model takes the number of distinct values $t$ of the booking horizon, minus 1 as the number of polynomials. To decrease the computational complexity, fewer polynomials could be fitted. In doing so, the impact on the overall performance of the model expressed in WAPE and revenue need to be researched. 
	
	The set-up to retrieve the results requires a 'warm-up period', for which t = 0 to t = 72 of the booking horizon is used. For this period, days similar to the forecasting date are selected and used. There is a potential revenue loss if RM optimisation isn't applied over this period. A simple strategy can tackle this warm-up period. This strategy can forecast if a given date is low, medium, or high demand. For each of these labels, a yield policy can be applied to counter the potential revenue loss.
	
	To further improve the cubic smoothing spline model, special events (eg. concerts or conferences) can be included. These events tend to occur more often during the weekend days than weekdays. Adding the events can give more context around the check-in dates. 
	
	\newpage
	
	\nocite{1, 2, 3, 4, 5, 7, 8, 9, 10, 11, 12, 13, 14, 15, 16, 17, 18, 19, 20}
	\bibliographystyle{plain}
	\bibliography{bibtex} 

\begin{thebibliography}{10}

\bibitem{12}
Lee{,} A.
\newblock {\em Airline reservations forecasting: probabilistic and statistical
  models of the booking process}.
\newblock Flight Transportation Laboratory Report{,} 232-236, 1990.

\bibitem{19}
Andrew{,} W.P.{,} Cranage{,} D.A.{,}~Lee{,} C.K.
\newblock {\em Forecasting hotel occupancy rates with time series models: An
  empirical analysis}.
\newblock Hospitality Research Journal{,} 14(2), 173-182, 1990.

\bibitem{10}
Santopietro{,} S.{,} Gargano{,} R.{,} Granata{,}~F.{,} de~Marinis{,}~G.
\newblock {\em Generation of Water Demand Time Series through Spline Curves}.
\newblock Journal of Water Resources Planning and Management{,}
  146(11):04020080, 2020.

\bibitem{5}
Claveria{,} O.{,}~Monte{,} E. and Torra{,} S.
\newblock {\em A new forecasting approach for the hospitality industry}.
\newblock International Journal of Contemporary Hospitality Management{,}
  27(7):1520-1538, 2015.

\bibitem{1}
Pizam{,}~A. (ed.).
\newblock {\em International Encyclopedia of Hospitality Management, sec. ed.}
\newblock Oxford: Butterworth Heinemann, 2010.

\bibitem{14}
Chu{,} F.L.
\newblock {\em Forecasting tourism demand with ARMA-based methods}.
\newblock Tourism Management{,} 30(5):740-751, 2009.

\bibitem{11}
Haensel{,} S.{,}~Koole{,} G.{,}.
\newblock {\em Estimating unconstrained demand rate functions using customer
  choice sets}.
\newblock Journal of Revenue and Pricing Management{,} 4(3):75-87, 2010.

\bibitem{2}
Mauri{,}~A. G.
\newblock {\em Hotel revenue management: Principles and practices}.
\newblock Milan: Pearson{,} 13:511-512, 2012.

\bibitem{16}
Talluri{,} K.{,}~Ryzin{,} G.
\newblock {\em Revenue management under a general discrete choice model of
  customer behavior}.
\newblock Management Science{,} 50(1):15-33, 2004a.

\bibitem{17}
Talluri{,} K.{,}~Ryzin{,} G.
\newblock {\em The Theory and Practice of Revenue Management}.
\newblock Springer, 2004b.

\bibitem{15}
Epperson{,} J.
\newblock {\em On the Runge Example}.
\newblock The American Mathematical Monthly{,} 94(4):329-341, 1987.

\bibitem{9}
Rich{,} J.
\newblock {\em A spline function class suitable for demand models}.
\newblock Econometrics and Statistics{,} 14, 2018.

\bibitem{13}
Shields{,} J.{,}~Shellemand{,} J.
\newblock {\em Small Business Seasonality: Characteristics and Management}.
\newblock Small Business Institute Journal{,} 9(1):37-50, 2013.

\bibitem{18}
Brown{,} L.D.{,} Zhang{,} R.{,}~Zhao{,} L.
\newblock {\em Root-Unroot Methods for Nonparametric Density Estimation and
  Poisson Random-Effects Models}.
\newblock Techical report{,} The Wharton School{,} Univ. Pennsylvania, 2002.

\bibitem{3}
Rajopadhye{,} M.{,}~Ghalia{,} M.B.{,} and Wang{,} P.P.
\newblock {\em Forecasting uncertain hotel room demand}.
\newblock Proceeding of the American Control Conference{,} 132(1-4):1-11, 1999.

\bibitem{20}
Loyola-González{,} O.
\newblock {\em Black-Box vs. White-Box: Understanding Their Advantages and
  Weaknesses From a Practical Point of View}.
\newblock IEEE Access{,} 7(1):154096-154113, 2019.

\bibitem{7}
Queenan{,} C.C.{,} Ferguson{,} M.{,} Higbie{,} J.{,}~Kapoor{,} R.
\newblock {\em A Comparison of Unconstraining Methods to Improve Revenue
  Management Systems}.
\newblock Production and operations management{,} 16(6), 2009.

\bibitem{4}
Weatherford{,}~L. R.{,} and Kimes{,}~S. E.
\newblock {\em A comparison of forecasting methods for hotel revenue
  management}.
\newblock International Journal of Forecasting{,} 19(3):401-415, 2003.

\bibitem{8}
Weatherford{,} L.R.{,}~Polt{,} S.
\newblock {\em Better unconstraining of airline demand data in revenue
  management systems for improved forecast accuracy and greater revenues}.
\newblock Journal of Revenue and Pricing Management{,} 1(3):234-254, 2002.

\end{thebibliography}
	
	\appendix
	\newpage
	\section{Lead Time Density by Day of Week} 
	\label{sec: Appendix - Lead Time density by DoW}
	
	Reservations made at $t$ of the booking horizon for stays on specific day of the week.
	
	\begin{figure}[H]
		\centering
		\includegraphics[width=.85\textwidth]{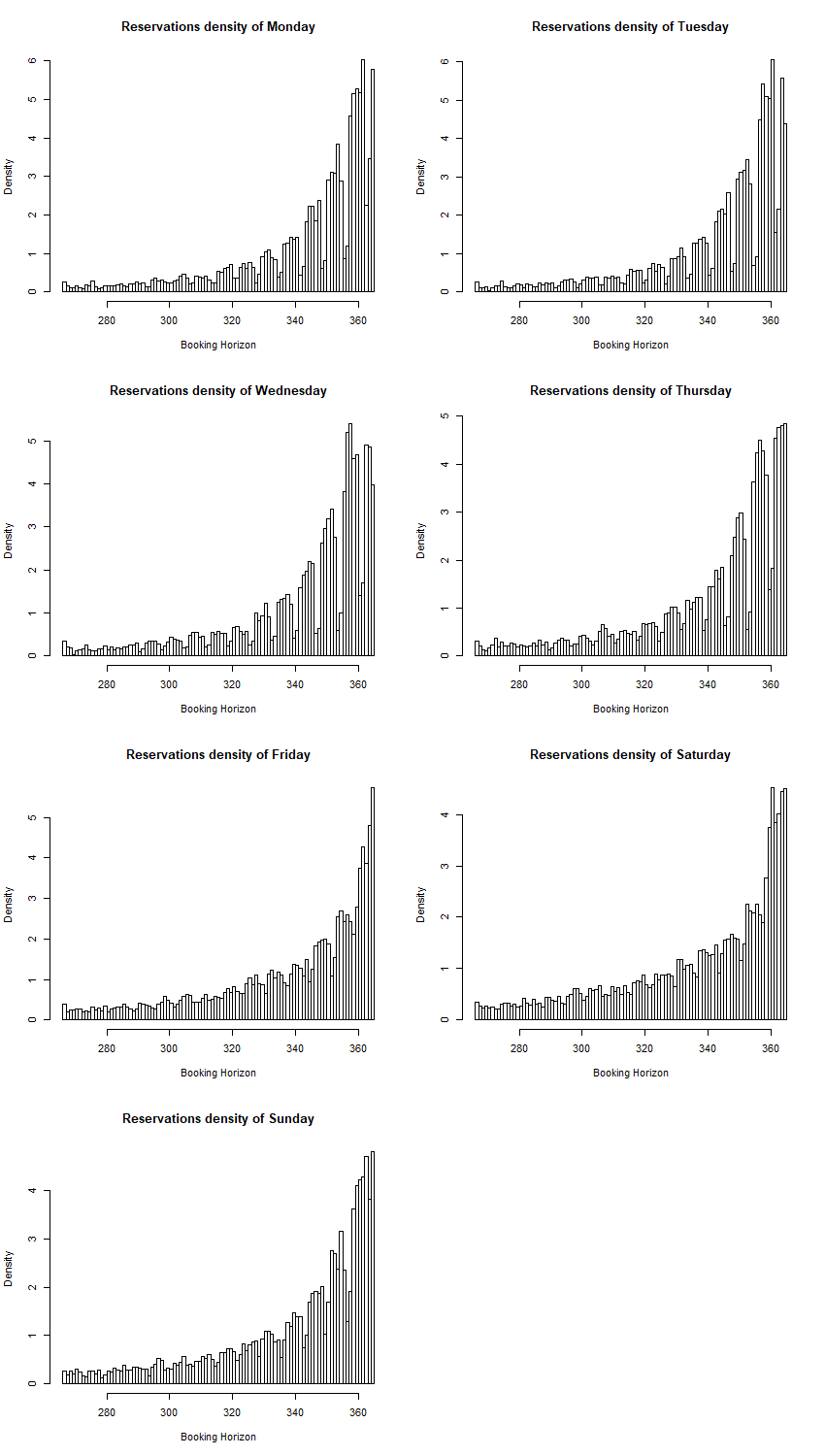}
		\caption{Density for lead time 265 - 365 per Day of Week}
		\label{fig:DA - Lead Time density by DoW}
	\end{figure}
	
	\newpage
	
	\section{Results per Hotel per Day of Week} 
	\label{sec: Appendix - Results WAPE}
	
	The tables in this section displays the WAPE per propery per day of week per customer choice set. The \textit{-} indicates no availability of data points in that customer choice set. 
	
	\begin{table}[H]
		\centering
		\addtolength{\leftskip} {-2cm}
		\addtolength{\rightskip}{-2cm}
		\begin{tabular}{l|lllllll}
			\textbf{}            & \multicolumn{7}{c}{\textbf{Hotel 1}}                                                                                                \\ \cline{2-8} 
			\textbf{Day of Week} & \textbf{Monday} & \textbf{Tuesday} & \textbf{Wednesday} & \textbf{Thursday} & \textbf{Friday} & \textbf{Saturday} & \textbf{Sunday} \\ \hline
			\textbf{Rate 70}     & 20.74           & 19.78            & 15.62              & 21.62             & 24.93           & 43.48             & 27.90           \\
			\textbf{Rate 80}     & 30.23           & 21.82            & 25.35              & 28.73             & 20.58           & 34.33             & 33.87           \\
			\textbf{Rate 90}     & 42.00           & 30.82            & 36.06              & 31.59             & 34.52           & 29.35             & 81.57           \\
			\textbf{Rate 100}    & 48.81           & 31.38            & 35.14              & 36.96             & 47.16           & 27.19             & 77.53           \\
			\textbf{Rate 110}    & 77.26           & 45.29            & 38.75              & 45.10             & 58.58           & 37.08             & 47.00           \\
			\textbf{Rate 120}    & 115.38          & 57.70            & 41.56              & 59.83             & 77.39           & 62.09             & 63.98           \\
			\textbf{Rate 130}    & 147.35          & 56.80            & 62.01              & 62.33             & 69.72           & 61.20             & 176.06          \\
			\textbf{Rate 140}    & 93.32           & 51.70            & 67.09              & 66.29             & 109.50          & 54.31             & 63.00           \\
			\textbf{Rate 150}    & 62.16           & 50.27            & 50.11              & 72.91             & 72.33           & 70.25             & 61.41           \\
			\textbf{Rate 160}    & 64.97           & 63.64            & 62.64              & 78.68             & 163.97          & 71.27             & 81.08           \\
			\textbf{Rate 170}    & 50.11           & 62.80            & 117.90             & 45.95             & 59.67           & 77.81             & 56.71          
		\end{tabular}
	   	\caption{Hotel 1 - WAPE per rate per day of week}
		\label{my-label-hotel-1}
	\end{table}

	\begin{table}[H]
		\centering
		\addtolength{\leftskip} {-2cm}
		\addtolength{\rightskip}{-2cm}
		\begin{tabular}{l|lllllll}
			\textbf{}            & \multicolumn{7}{c}{\textbf{Hotel 2}}                                                                                                \\ \cline{2-8} 
			\textbf{Day of Week} & \textbf{Monday} & \textbf{Tuesday} & \textbf{Wednesday} & \textbf{Thursday} & \textbf{Friday} & \textbf{Saturday} & \textbf{Sunday} \\ \hline
			\textbf{Rate 100}    & 29.98   & 169.69  & 16.56     & 15.69    & 13.21   & 35.12    & 19.39   \\
			\textbf{Rate 115}    & 19.60   & 19.05   & 14.95     & 21.29    & 15.71   & 17.30    & 28.51   \\
			\textbf{Rate 130}    & 21.52   & 22.35   & 26.64     & 20.15    & 38.10   & 19.11    & 90.64   \\
			\textbf{Rate 145}    & 22.67   & 22.92   & 21.88     & 24.59    & 57.11   & 16.19    & 88.27   \\
			\textbf{Rate 160}    & 43.08   & 29.24   & 30.68     & 25.58    & 54.41   & 24.81    & 80.00   \\
			\textbf{Rate 175}    & 56.79   & 24.21   & 29.15     & 35.37    & 98.36   & 36.37    & 74.99   \\
			\textbf{Rate 190}    & 93.91   & 30.17   & 33.63     & 97.33    & 66.62   & 47.39    & 50.00   \\
			\textbf{Rate 205}    & 98.16   & 44.23   & 40.39     & 65.84    & 75.00   & 81.30    & 20.00   \\
			\textbf{Rate 220}    & 29.81   & 41.10   & 49.29     & 58.00    & 83.05   & 240.91   & -      \\
			\textbf{Rate 235}    & 88.34   & 60.80   & 50.73     & 78.78    & 78.86   & 234.44   & -      \\
			\textbf{Rate 250}    & 400.00  & 85.17   & 53.90     & 54.00    & 77.03   & 327.24   & -         
		\end{tabular}
		\caption{Hotel 2 - WAPE per rate per day of week}
		\label{my-label-hotel-2}
	\end{table}

	\begin{table}[H]
		\centering
		\addtolength{\leftskip} {-2cm}
		\addtolength{\rightskip}{-2cm}
		\begin{tabular}{l|lllllll}
			\textbf{}            & \multicolumn{7}{c}{\textbf{Hotel 3}}                                                                                                \\ \cline{2-8} 
			\textbf{Day of Week} & \textbf{Monday} & \textbf{Tuesday} & \textbf{Wednesday} & \textbf{Thursday} & \textbf{Friday} & \textbf{Saturday} & \textbf{Sunday} \\ \hline
			\textbf{Rate 90}  & 13.59   & 20.41   & 26.64     & 19.03    & 24.66   & 20.16    & 18.95   \\
			\textbf{Rate 105} & 36.95   & 34.97   & 32.87     & 29.69    & 18.53   & 26.03    & 33.20   \\
			\textbf{Rate 120} & 46.23   & 34.90   & 32.55     & 31.19    & 28.77   & 29.36    & 35.20   \\
			\textbf{Rate 135} & 68.47   & 30.92   & 38.65     & 29.25    & 33.30   & 23.82    & 39.10   \\
			\textbf{Rate 150} & 58.56   & 54.43   & 65.75     & 48.20    & 50.84   & 30.62    & 45.67   \\
			\textbf{Rate 165} & 85.61   & 55.17   & 60.07     & 49.48    & 56.75   & 51.36    & 78.00   \\
			\textbf{Rate 180} & 132.28  & 61.02   & 83.50     & 52.48    & 50.42   & 89.83    & 69.58   \\
			\textbf{Rate 195} & 144.91  & 51.46   & 61.34     & 93.25    & 58.52   & 80.19    & 88.50   \\
			\textbf{Rate 210} & 237.80  & 58.92   & 82.08     & 346.07   & 96.91   & 143.16   & 58.38   \\
			\textbf{Rate 225} & 122.29  & 111.20  & 83.40     & 183.54   & 130.81  & 147.50   & 123.56  \\
			\textbf{Rate 240} & 115.60  & 141.94  & 56.05     & 86.56    & 72.30   & 62.82    & 99.51  
		\end{tabular}
		\caption{Hotel 3 - WAPE per rate per day of week}
		\label{my-label-hotel-3}
	\end{table}

	\begin{table}[H]
		\centering
		\addtolength{\leftskip} {-2cm}
		\addtolength{\rightskip}{-2cm}
		\begin{tabular}{l|lllllll}
			\textbf{}            & \multicolumn{7}{c}{\textbf{Hotel 4}}                                                                                                \\ \cline{2-8} 
			\textbf{Day of Week} & \textbf{Monday} & \textbf{Tuesday} & \textbf{Wednesday} & \textbf{Thursday} & \textbf{Friday} & \textbf{Saturday} & \textbf{Sunday} \\ \hline
				\textbf{Rate 150} & 19.46   & 15.04   & 16.71     & 13.59    & 20.10   & 27.61    & 19.83   \\
				\textbf{Rate 170} & 31.41   & 25.02   & 25.93     & 24.57    & 30.86   & 34.47    & 30.23   \\
				\textbf{Rate 190} & 34.51   & 20.79   & 44.42     & 22.30    & 22.41   & 23.52    & 36.03   \\
				\textbf{Rate 210} & 27.77   & 29.31   & 37.77     & 25.81    & 18.79   & 26.38    & 93.22   \\
				\textbf{Rate 230} & 26.91   & 29.59   & 20.89     & 26.66    & 25.83   & 30.35    & 196.77  \\
				\textbf{Rate 250} & 31.25   & 23.20   & 20.48     & 32.88    & 30.41   & 34.47    & 64.69   \\
				\textbf{Rate 270} & 48.06   & 30.80   & 29.00     & 37.92    & 44.77   & 27.12    & 42.17   \\
				\textbf{Rate 290} & 56.46   & 34.08   & 28.31     & 44.02    & 52.49   & 31.67    & 67.00   \\
				\textbf{Rate 310} & 50.70   & 43.13   & 39.16     & 49.31    & 53.06   & 38.37    & 61.00   \\
				\textbf{Rate 330} & 73.80   & 52.48   & 55.47     & 52.36    & 43.42   & 51.11    & 56.00   \\
				\textbf{Rate 350} & 47.15   & 71.16   & 54.15     & 77.87    & 59.83   & 59.53    & 53.16   \\
				\textbf{Rate 370} & 57.41   & 61.25   & 70.16     & 96.96    & 56.74   & 67.27    & 78.83   \\
				\textbf{Rate 390} & 79.57   & 49.21   & 89.89     & 51.12    & 60.46   & 80.40    & 92.00   \\
				\textbf{Rate 410} & 53.91   & 71.24   & 69.25     & 32.29    & 78.33   & 61.74    & -      \\
				\textbf{Rate 430} & 89.19   & 339.00  & 83.38     & 109.00   & 41.62   & 70.98    & -      \\
				\textbf{Rate 450} & 56.57   & 117.00  & 67.86     & 400.00   & 31.43   & 26.07    & -     
		\end{tabular}
		\caption{Hotel 4 - WAPE per rate per day of week}
		\label{my-label-hotel-4}
	\end{table}
	\newpage
		
\end{document}